# Flexible but Refractory Single-Crystalline Hyperbolic Metamaterials

Ruyi Zhang,*,[†,‡] Ting Lin,[§] Shaoqin Peng,[†,‡] Jiachang Bi,[†,‡] Shunda Zhang,[†,‡] Guanhua Su,[†,‡] Jie Sun,[†] Junhua Gao,[†] Hongtao Cao,[†] Qinghua Zhang,*,[§] Lin Gu,[§] and Yanwei Cao*,[†,‡]

[†]Ningbo Institute of Materials Technology and Engineering, Chinese Academy of Sciences, Ningbo 315201, China
[‡]Center of Materials Science and Optoelectronics Engineering, University of Chinese Academy of Sciences, Beijing 100049, China
[§]Beijing National Laboratory for Condensed Matter Physics, Institute of Physics, Chinese Academy of Sciences, Beijing 100190, China

**ABSTRACT:** The fabrication of flexible single-crystalline plasmonic or photonic components in a scalable way is fundamentally important to flexible electronic and photonic devices with high speed, high energy efficiency, and high reliability. However, it remains to be a big challenge so far. Here, we have successfully synthesized flexible single-crystalline optical hyperbolic metamaterials by directly depositing refractory nitride superlattices on flexible fluorophlogopite-mica substrates with magnetron sputtering. Interestingly, these flexible hyperbolic metamaterials show dual-band hyperbolic dispersion of dielectric constants with low dielectric losses and high figure-of-merit in the visible to near-infrared ranges. More importantly, the optical properties of these nitride-based flexible hyperbolic metamaterials show remarkable stability under either 1000 °C heating or 1000 times bending. Therefore, the strategy developed in this work offers an easy and scalable route to fabricate flexible, high-performance, and refractory plasmonic or photonic components, which can significantly expand the applications of current electronic and photonic devices.

For the pursuit of high-performance electronic and optical devices with high speed, high energy efficiency, and high reliability, the study of single-crystalline functional materials has attracted tremendous attention recently.[1-4] Generally, single-crystalline functional materials and devices are prepared on planar and rigid substrates. However, with the continuously growing demands for devices to be applied in smart, lightweight, and wearable scenarios, there emerges great interest in flexible functional materials.[5-9] Despite significant progress has been made in the design, fabrication, and process of flexible functional materials and devices in the past few years,[10-13] the scalable fabrication of flexible single-crystalline electronic/photonic materials and devices is still very challenging,[14] especially for the plasmonic materials and devices.

Plasmon, the collective excitation of conduction electrons, has revolutionized the way of light-electron-matter interactions in functional materials,[15] leading to many important applications in electronics, nanophotonics, and electro/photo-catalysis.[16-20] Integrating plasmonic/photonic components with flexible materials and exploiting their unique advantages (such as mechano-photonic coupling and geometric degrees of freedom) have already been the research frontiers of plasmonics and nanophotonics.[21-25] At present, most plasmonic/photonic materials (e.g., noble metals Au and Ag, semiconductors, transparent conducting oxides, metal nitrides, and two-dimensional materials) are inorganic materials, which are incompatible with conventional flexible organic materials in synthesis temperature, crystal structure, thermal expansion coefficients, and to mention a few.[23] Therefore, it is extremely challenging to directly grow single-crystalline plasmonic/photonic materials on flexible organic substrates. As an important paradigm of plasmonic/photonic components, hyperbolic metamaterials (HMMs) with extreme optical anisotropy rarely existed in nature have been applied in the fields of negative refraction,[26] subwavelength-resolution imaging,[27] highly sensitive sensors,[28] and spontaneous emission engineering[29] lately. Single-crystalline HMMs can be directly grown on rigid substrates.[30-33] However, their flexible counterparts haven't been successfully fabricated yet. It is also important to note that typical organic substrates are extremely vulnerable to harsh environments, e.g., high temperature or intensive light illumination,[23-25] which strongly limit the applications of flexible plasmonic/photonic components. Therefore, developing a strategy for scalable fabrication of flexible single-crystalline plasmonic/photonic components with high performance and high-temperature stability is very important, which can significantly expand their applications.

Here, we demonstrate a new fabrication method for flexible high-performance plasmonic/photonic components by directly depositing single-crystalline HMMs on flexible inorganic fluorophlogopite-mica (F-mica). These flexible HMMs were prepared by synthesizing transition-metal nitride (TMN) superlattices (SLs) on F-mica via reactive magnetron sputtering, a scalable deposition technology. These flexible TMN SLs show low-loss hyperbolic dispersion of dielectric constants and mechanical robustness upon repeated bending. It is remarkable that owing to the excellent thermal stabilities of TMN SLs and F-mica substrates, these flexible HMMs can also be refractory and applied in extreme environments. This study provides a novel, easy, and scalable route to fabricate flexible, high-performance, refractory plasmonic/photonic components, which can significantly expand the application of current flexible plasmonic/photonic devices.

As shown in Figure 1a, HMMs composed of multilayered single-crystalline titanium nitride (TiN)/scandium nitride (ScN) SLs were directly deposited on F-mica substrates by a homemade reactive magnetron sputtering system with 2-inch pure titanium (99.999%) and scandium (99.995%) targets and pure nitrogen (99.999%) reactive gas. These TMN SLs can be highly flexible through simple mechanical exfoliation of F-mica substrates. It is noteworthy that both TiN and ScN are rock salt-structured refractory materials (lattice parameters of $a = 4.24$ Å and $a = 4.50$ Å) with extremely high melting temperatures of 2930°C and 2600°C, respectively.[34] TiN is a widely applied material in hard coating, plasmonic, superconducting, and semiconductor devices due to its remarkable mechanical hardness, excellent plasmonic performance comparable to noble metals (such as Au), large kinetic inductance in superconducting states, and superior complementary-metal-oxide semiconductor (CMOS) compatibility.[35-36] On the other hand, ScN is an indirect bandgap (0.9 eV) semiconductor with variable applications in thermoelectrics, plasmonics, and thermophotonics,[37] which can act as a dielectric layer in visible and near-infrared (*vis*-NIR) ranges here when forming HMMs with metallic TiN layer.

As shown in Figure 1b, single-crystalline F-mica [$KMg_3Al(Si_3O_{10})F_2$] substrate is a synthetic two-dimensional van der Waals layered material with lattice parameters $a$ = 5.31 Å, $b$ = 9.18 Å, $c$ = 10.14 Å, $\alpha$ = 90°, $\beta$ = 100.07°, $\gamma$ = 90°.[38] There are several benefits to choosing F-mica as a substrate for flexible plasmonic/photonic components. Firstly, it shows high transparency (over 90%) and low loss in *vis*-NIR ranges (see optical properties of F-mica substrate in Figure S1), allowing optical applications in a wide spectrum. Secondly, it is chemically inert, thermally stable up to 1100 °C, and atomically flat on the surface,[39-40] which are very appealing for device patterning, high-temperature heat treatments, and practical applications in extreme environments. Lastly, mica is mechanically robust as a flexible substrate upon repeated bending.[40-45] Therefore, F-mica can be an ideal candidate as a substrate for the high-performance, flexible, refractory plasmonic/photonic components. Moreover, as illustrated in Figure 1c, (001)-F-mica possesses a pseudohexagonal lattice for its layered planes with a lattice parameter ~ 5.31 Å, which is close to the lattice parameters (~ 5.19 Å and 5.51 Å) of the hexagonal lattices of TiN and ScN supercells. Therefore, F-mica can be applied in the epitaxial growth of high-quality rock-salt TMN films, e.g., TiN and ScN, by following the epitaxial relationship expressed as TMN[11-2]∥F-mica[100] and TMN[1-10]∥F-mica[010] for the in-plane directions and TMN(111)∥F-mica(001) for the out-of-plane direction.[36]

Two TiN/ScN SLs, [TiN(50 Å)/ScN(42 Å)]$_{10}$ and [TiN(50 Å)/ScN(76 Å)]$_8$ (denoted as SL1 and SL2, respectively), were deposited on F-mica. As shown in Figure 1d, the (111) diffraction peaks with satellite peaks for SLs can be seen around the shoulders of F-mica (004) diffraction peaks in the 2$\theta$-$\omega$ scans, indicating (111)-oriented growth of SLs without detectable secondary phases (also see wide-range 2$\theta$-$\omega$ scans in Figure S2). Due to distinct differences in lattice parameters between ScN and TiN (as indicated by their bulk diffraction peak positions), the diffraction patterns of SLs are composed of 0-order ScN and 0-order TiN diffraction peaks with their high-order satellite peaks. According to the spacings of these satellite peaks in the 2$\theta$-$\omega$ scans, the period thicknesses of ~ 95 Å and ~130 Å can be estimated for SL1 and SL2, respectively, agreeing very well with the extracted period thicknesses of 92 Å and 126 Å from the fitted X-ray reflectivity curves (see Figure S3). The epitaxy of TiN/ScN SLs on F-mica was further verified by performing $\varphi$ scans around SL (002) and F-mica (202) diffractions in Figure S4a. As seen, the 6-fold symmetry rather than the 3-fold symmetry for the SLs indicates a twin structure of SLs, which is commonly seen in epitaxial films on mica (001) and $Al_2O_3$ (0001) substrates.[41-42, 46-47] The $\varphi$ scans can also present the expected epitaxial relationship (despite the twin feature) for TiN/ScN SL on F-mica. The reciprocal space mapping around F-mica (-207) diffraction (see Figure S4b) shows mosaic spreading for the SL diffractions, indicating the presence of lattice misfit-induced dislocations, which have been reported in other epitaxial films grown on mica.[40-45] We also deposited TiN(50 Å)/ScN(42 Å)]$_{10}$ on a 2 in. Al$_2$O$_3$ (0001) wafer to compare the structural and optical properties of flexible TiN/ScN SL with those on the rigid substrate (Figure S5).

Next, to investigate the crystal structures of TiN/ScN SLs on F-mica at the atomic scale, scanning transmission electron microscope (STEM) characterization was carried out. As seen in Figure 2a, all selected-area-electron-diffractions (SAED) images collected near the surface (region I), the interface between F-mica and superlattice (region II), and the substrate (region III) of SL1 (regions I-IV are also shown in Figure 2b for clarity) show clear single-crystalline patterns, further verifying the epitaxial relationships expressed as, ScN[11-2]∥TiN[11-2]∥F-mica[100] and ScN[1-10]∥TiN[1-10]∥F-mica[010] for the in-plane directions and ScN(111)∥TiN(111)∥F-mica(001) for the out-of-plane direction. The horizontal separation of ScN and TiN SAED patterns of region I in Figure 2a reveals that the ScN layer and TiN layer are not coherently strained, which is consistent with the results from the RSM data. Moreover, the low-magnification STEM images and corresponding energy-dispersive X-ray (EDX) mappings in Figure 2b,c indicate the periodic alternating stacks of TiN and ScN layers, the smooth interfaces between TiN and F-mica, and the smooth interfaces between ScN and TiN layers for SL1 and SL2. Figure 2d,e show the high-angle annular dark-field (HAADF)-STEM images acquired near the mica-superlattice interface area (region IV) of SL1 and surface area (region V) of SL2, respectively. Both HAADF-STEM images are viewed along the F-mica[010] and ScN/TiN[1-10] zone axis, which displays (111)-oriented atomic arrangements of Ti and Sc elements on (001)-oriented F-mica. It is worth noting that the epitaxy of TMN layer with a smooth interface/surface is critical to its high plasmonic performance and low dielectric loss characteristic.[48] According to the atomic-level analysis from STEM, it is remarkable that the epitaxy with smooth interfaces extends throughout the whole superlattices on F-mica despite of relatively large lattice misfit (~ 6 %) between ScN and TiN. Nevertheless, the developed method for epitaxial superlattice with ultra-thin (< 10 nm) and smooth TMN layers on flexible F-mica has overcome a significant challenge of direct fabrication of single-crystalline plasmonic/photonic components on flexible substrates in the previous study. Moreover, this method should also be suitable for many other rock salt nitride layers with closed lattice parameters (4.2 - 4.6 Å), e.g., ZrN, HfN, VN, etc.

Then, the successful epitaxy of TMN SLs on F-mica here enables an investigation into their performance as flexible HMMs and a comparison with other HMMs. The optical properties of TiN/ScN SLs on F-mica were characterized by a spectroscopic ellipsometer (SE). Since the thicknesses of both TiN and ScN layers in SLs are much smaller than the light wavelength, the uniaxial anisotropy models can be applied to deduce the dielectric constants in the orientations parallel ($\varepsilon_\parallel$) and perpendicular ($\varepsilon_\perp$) to the SL plane from the measured SE data. It is interesting to note that the optical properties of both TiN films (can be referred to our previous work[36]) and ScN films (Figure S6) on F-mica substrates are thickness-dependent. As seen in Figure S7, the SE data for SL1 and SL2 can be well fitted with mean-squared error (MSE) around 6. Consistent with other metal nitride SLs on rigid substrates (e.g., MgO),[30-32] the TiN/ScN SLs on F-mica here also show two different hyperbolic regions with the real parts of dielectric constants ($\varepsilon'_\perp$ and $\varepsilon'_\parallel$) showing the opposite signs, i.e., type-I HMM ($\varepsilon'_\perp < 0, \varepsilon'_\parallel > 0$) in visible ranges and type-II HMM ($\varepsilon'_\perp > 0, \varepsilon'_\parallel < 0$) in *vis*-NIR ranges (Figure 3a-c). As seen, SL1 and SL2 are type-I HMM in the wavelength ranges 496-534 nm and 496-583 nm with very low dielectric losses ($\varepsilon''_\parallel < 2$ and $\varepsilon''_\perp < 11$), which are very promising for hyperlens applications.[31] Moreover, both SL1 and SL2 are also type-II HMM above wavelengths 641 nm and 750 nm with dielectric loss below 15 and 9, respectively, which are very useful for the photonic density of state engineering applications.[31] Due to the low dielectric losses, transmittances of 25% and 37 % at ~ 550 nm can still be seen for 92 nm-thick SL1 and 101 nm-thick SL2 (Figure S8), respectively. On the other hand, it is interesting to compare the hyperbolic behaviors of TiN/ScN SL on flexible F-mica with those on rigid $Al_2O_3$ (0001) wafers (Figure S5). As seen, the [TiN(50 Å)/ScN(42 Å)]$_{10}$ SL on 2 in. $Al_2O_3$ (0001) wafer shows similar SL (111) diffractions, hyperbolic regions (type-I HMM in 499-544 nm ranges and type-II HMM above 644 nm), and low dielectric losses as those for SL1 on F-mica, indicating that the SL grown on flexible F-mica substrates and rigid $Al_2O_3$ (0001) wafers almost have the same crystalline structures and optical performance. To gain deeper insight into the optical anisotropy in TiN/ScN SLs on F-mica, both experimental (Exp.) and fitted (Fit.) (deduced from the same fitting models used in Figure 3b,c) s-polarized and p-polarized reflectivities are presented in Figure 3d-i. It can be seen that the fitted s- and p-polarized reflectivities agree well with the experimental reflectivities for both SL1 and SL2, validating the well-deduced dielectric constants by the same fitting models in Figure 3b,c. At a fixed incidence angle, the s-polarized reflectivities are much higher than the p-polarized reflectivities, indicating the optical anisotropy of SL on F-mica. With increasing the incidence angles, the s-polarized reflectivities increase while the p-polarized reflectivities decrease in NIR ranges. The Brewster's angles, where the p-polarized reflectivities reach zero at visible wavelengths, are satisfied at an incidence angle of ~ 65° for SL1 and ~ 55° for SL2, respectively.

To further evaluate the performance of HMMs in this work, the figure of merits (FOMs) defined as $Re(k_\perp)/Im(k_\perp)$,[30, 48] where $Re(k_\perp)$ and $Im(k_\perp)$ are the real and imaginary part of the propagation components perpendicular to the film plane, were calculated for SL1 and SL2, respectively. Due to the scarcity of flexible high-quality SLs, the FOMs of SL1 and SL2 are then compared with other multilayered HMMs systems on rigid substrates (denoted by *) with filling factors of ~ 50 % for the metallic layer. As seen in Figure 4, the FOMs of flexible HMMs in this work surpass the FOMs of reported HMMs in the wavelength ranges 430 - 570 nm for SL1 and 450 - 750 nm for SL2. Above ~ 750 nm, the TiN/(Al,Sc)N* SL on MgO substrate with very small lattice misfits (< 1 %) between TiN, (Al,Sc)N, and MgO has very few misfit-induced dislocations and the highest FOM.[30] Despite being lower than the FOMs (e.g., ~ 0.5 at 1500 nm) of TiN/(Al,Sc)N* SL and ZrN/ScN* SL above ~ 750 nm,[30, 32] the FOMs (e.g., ~ 0.18 at 1500 nm) of SL1 and SL2 are very close to the FOM of HfN/ScN* SL in previous work.[32] It is also important to note that the multilayers based on noble metals (e.g., Au and Ag) and dielectric oxides (e.g., $TiO_2$ and $Al_2O_3$) are the most widely applied HMMs for the applications of hyperlens, spontaneous emission engineering, negative refraction, and nonlinear optics.[27, 29] Comparing to these metal/dielectric oxide-based HMMs (e.g., FOMs of 0.01-0.05 at 1500 nm), our flexible HMMs based on single-crystalline nitride SLs show much higher performance above ~ 500 nm.

To reveal the mechanical flexibility and thermal stability, we conducted further tests on TiN/ScN SLs-based HMMs (Figure 5). The TiN/ScN SLs on F-mica can be highly flexible after mechanical exfoliation (the thickness of F-mica was reduced to ~ 5 μm). To perform the bending test, the TiN/ScN SLs were transferred onto a blue tape and mounted on a displacement actuator using a stepper motor. As seen in Figure 5a, the TiN/ScN SLs can be gradually bent from a flat state to concave curvature with a minimum bending radius of ~ 2 mm. The maximum in-plane strain applied to TiN/ScN SLs during bending is around 0.125 %, which is much smaller than the typical critical strain of fracture ($\geq$ 1.18 %) for monolithic TiN films or SLs (see calculations for the strain applied to SLs in Supporting Information). Moreover, a comparison of STEM images between the TiN/ScN SLs before and after repeated bending indicates that repeated bending doesn't induce cracks in SLs around misfit-induced dislocations (Figure S9). Therefore, the TiN/ScN SLs are mechanically robust and highly flexible upon repeated bending. The TiN/ScN SLs can also be wrapped around a cylindrical stick (Figure S10), showing the potential in rollable device application.To further investigate the effect of high-temperature annealing and repeated bending on the optical properties of TiN/ScN SLs, the testing procedures as depicted in Figure 5b were applied to SL2 sample. Firstly, the as-grown SL2 sample, denoted as the pristine sample, was annealed in the high vacuum (~ $10^{-7}$ Torr) at 1000 °C for 12 h. Then, the mechanical exfoliation was conducted before bending the thinned sample ~ 1000 times. After each procedure, the variable-angle SE characterization was carried out immediately. As shown in Figure S11, post-annealing and repeated bending tests cause very slight changes to the original SE data of SL2 sample. The dielectric constants extracted from these SE data in Figure 5c reveal that post-annealing and repeated bending only lead to very slight drift (< 5 nm) of type-I and type-II hyperbolic regions. No significant changes in the dielectric losses were observed after post-annealing and repeated bending. The above experiments indicate that the TiN/ScN SLs can remain highly flexible and stable after high-temperature treatments. It is interesting to further compare our flexible HMMs based on TiN/ScN SLs with those existing flexible multilayered HMMs in terms of crystallinity of functional components, HMM ranges, operating temperature, etc. As seen in Table 1, the functional components of flexible multilayered HMMs in previous reports mainly consist of dozens nanometer thick- non-epitaxial Au and organic layers, such as poly(vinyl alcohol) (PVA) and poly(methyl methacrylate) (PMMA).[49-50] Since typical noble metals have high surface energy, it is difficult to deposit ultrathin (< 10 nm) noble metal films with very smooth surface morphology. Therefore, it would be very challenging to further shrink the optical devices and lower the optical loss due to surface roughness. Those drawbacks can be overcome by applying TMN functional layers with moderate surface energy. Through successful epitaxy, the interfaces of TMN SLs are smooth at the nanoscale, which is helpful for lower optical losses and higher figures of merits. More importantly, as both functional components (TMN SLs) and F-mica are thermally stable at high temperatures, the operating (heat-treatment) temperature can be significantly improved to ~ 1000 ℃ in this work, which is much higher than the operating temperatures (~ 100 ℃) of existing flexible multilayered HMMs with organic materials. Therefore, the refractory property shown by TiN/ScN SLs can significantly broaden the application of current flexible HMMs in harsh environments.

In summary, epitaxial superlattices consisting of ultrathin (< 10 nm) TMN (e.g., ScN and TiN) layers have been successfully deposited on flexible F-mica substrates using reactive magnetron sputtering. These flexible TiN/ScN superlattices show low-loss HMM behaviors and high figure of merits in *vis*-NIR ranges, which can even compete with those prepared on rigid substrates. Moreover, these flexible TiN/ScN SLs-based HMMs show remarkable stability even under high-temperature treatment (1000 ℃) and repeated bending (1000 times). These flexible TiN/ScN SLs with excellent HMM performance, thermal stability, and mechanical flexibility in this work outperform those flexible HMMs consisting of noble metal and dielectrics in previous reports. Our study provides a novel, easy, and scalable route to synthesizing flexible, high-performance, and refractory single-crystalline plasmonic/photonic components, which can significantly increase stability and expand their application in harsh environments.

## ASSOCIATED CONTENT

### Supporting Information

Supporting Information is available online.
Experimental details, wide range $2\theta$-$\omega$ scans, $\varphi$ scans, and reciprocal space mapping for TiN/ScN SL on F-mica (001), photo image, structural and optical properties for TiN/ScN SL on 2 in. $Al_2O_3$ (0001) wafer, optical properties of F-mica and ScN films, fitting details of SE data, UV-*vis*-NIR transmittance, STEM images of SLs after repeated bending, SE data for thermal stability and flexibility tests.

## AUTHOR INFORMATION

### Corresponding Author

*E-mail: zhangruyi@nimte.ac.cn, zqh@iphy.ac.cn, ywcao@nimte.ac.cn

### Notes

The authors declare no competing financial interest.

**Author Contributions**

R. Zhang and T. Lin contributed equally to this work. R. Zhang and Y. Cao conceived the project. R. Zhang, S. Peng, J. Bi, S. Zhang, G. Su, and J. Sun prepared the samples and performed the HRXRD and SE measurements. T. Lin, L. Gu, and Q. Zhang performed the STEM characterizations and analyzed the STEM images. R. Zhang, J. Gao, and H. Cao conducted the fittings and analyses of SE data. All authors discussed the data and contributed to the manuscript.

ACKNOWLEDGMENTS

This work was supported by the National Key R&D Program of China (Grant No. 2022YFA1403000), the National Natural Science Foundation of China (Grant Nos. 11874058, U2032126, 52072400, and 52025025), the Pioneer Hundred Talents Program of the Chinese Academy of Sciences, the Beijing Natural Science Foundation (Z190010), the Natural Science Foundation of Zhejiang Province (Grant No. LXR22E020001), the Beijing National Laboratory for Condensed Matter Physics, and the Ningbo Science and Technology Bureau (Grant No. 2022Z086). This work was partially supported by the Youth Program of the National Natural Science Foundation of China (Grant No. 12004399).

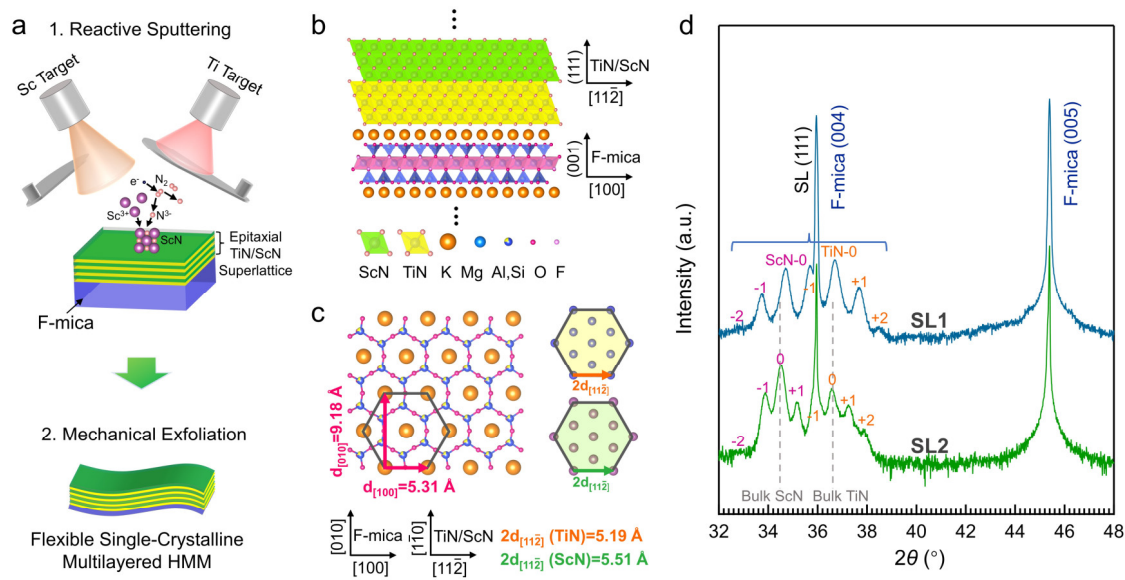

**Figure 1.** (a) Schematic process of synthesizing flexible, single-crystalline, hyperbolic metamaterials through direct epitaxy of transition-metal nitride superlattices on F-mica substrates and mechanical exfoliation. (b) Side view and (c) top view of crystal structures of TiN/ScN superlattices on F-mica. (d) $2\theta$-$\omega$ scans of TiN/ScN superlattices (SL1 and SL2) around F-mica (004) and (005) diffractions.

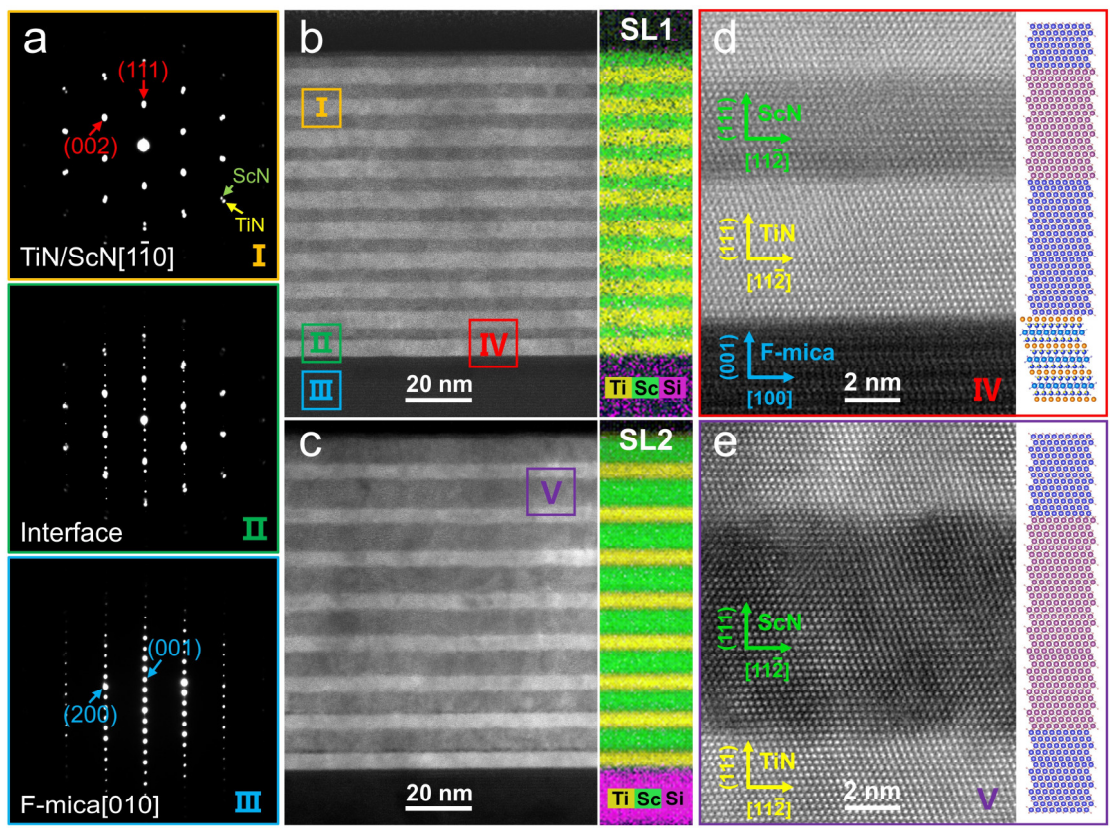

**Figure 2.** STEM characterizations of TiN/ScN superlattices (SL1 and SL2) on F-mica. (a) SAED images of SL1 near the surface (region I), interface (region II), and substrate (region III). (b) and (c) The low-magnification STEM images and corresponding EDX mappings of SL1 and SL2, respectively. (d) The HAADF-STEM image around the interface of SL1 (see the region IV in (b)) viewed along F-mica [010] zone axis. (e) The HAADF-STEM image near the surface of SL2 (see the region V in (c)) viewed along TiN and ScN [1-10] zone axis. The regions I to V are denoted as squares in (b) and (c) for clarity.

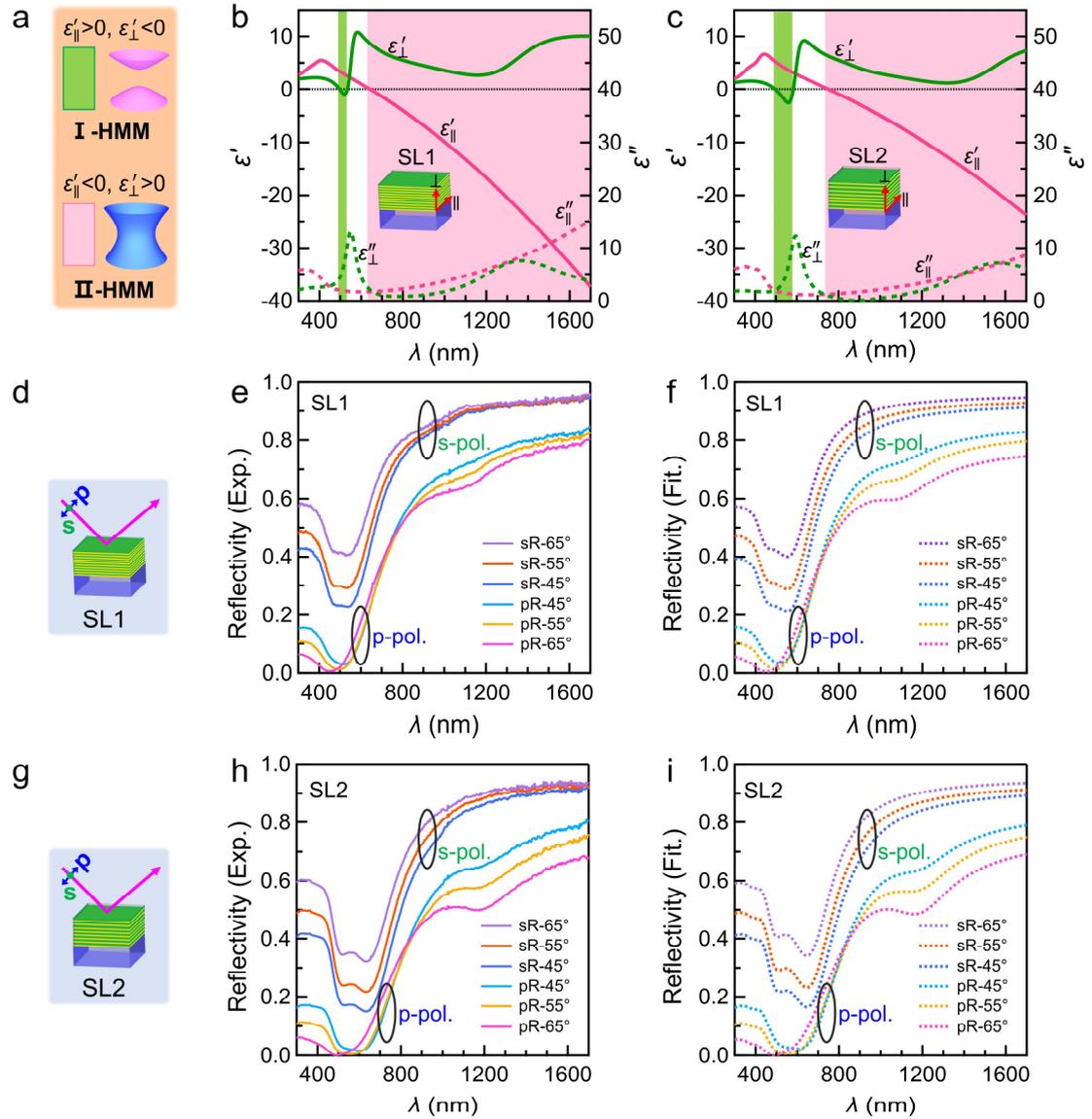

**Figure 3.** Optical properties of TiN/ScN superlattices. (a) The definition of type-I (green region) and type-II (pink region) hyperbolic dispersion of dielectric constants. The dielectric constants in the directions parallel (∥) and perpendicular (⊥) to the SL plane for SL1 and SL2 are shown in (b) and (c), respectively. The optical configuration for polarized reflectivity, experimental (Exp.) and fitted (Fit.) s-polarized (s-pol.) and p-polarized (p-pol.) reflectivities (sR and pR) at incidence angles 45°, 55°, and 65° are shown in (d), (e), (f) for SL1 sample and in (g), (h), (i) for SL2 sample, respectively.

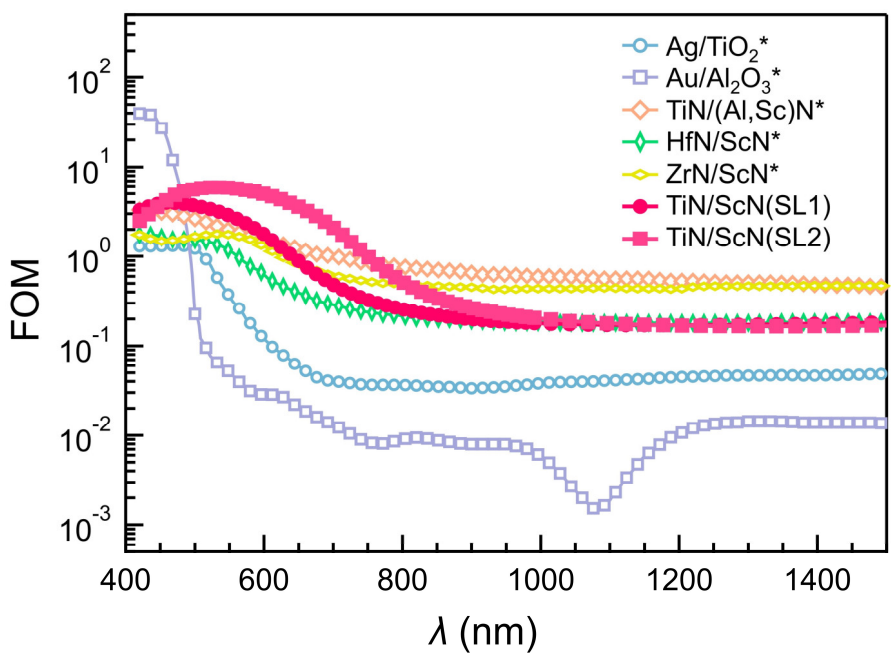

**Figure 4.** Summary of the figure of merits (FOMs) for flexible HMMs (TiN/ScN SLs on F-mica) in this work and FOMs of multilayered HMMs on rigid substrates in previous reports (denoted as *).[32] (Reprinted in part with permission from ref. 32. Copyright 2022 American Chemical Society.)

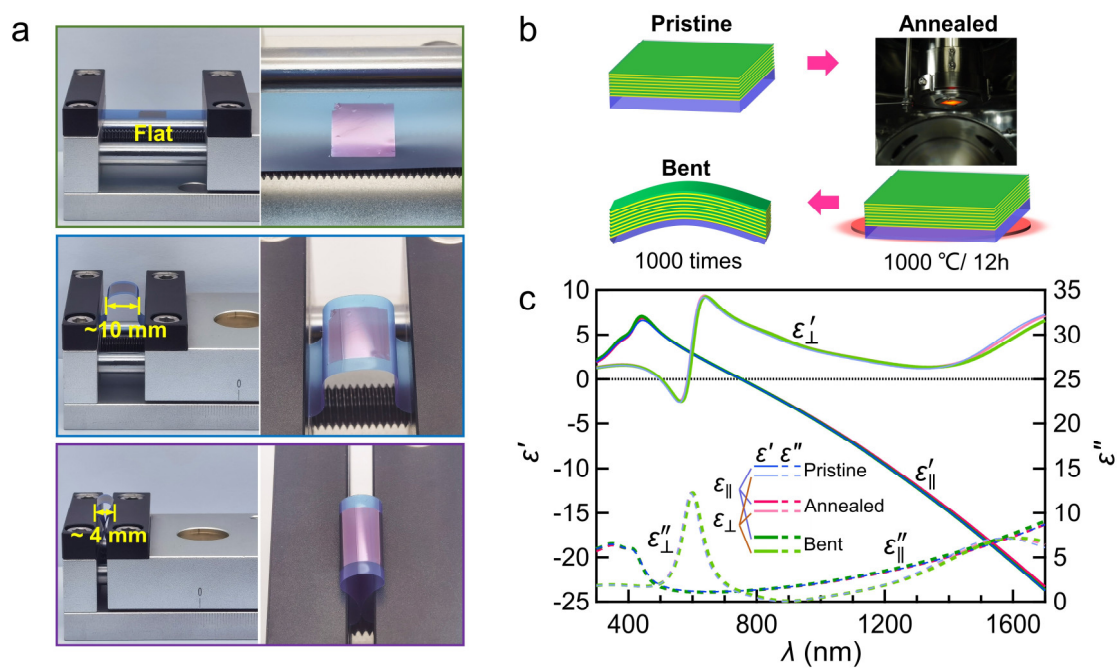

**Figure 5.** Flexibility and thermal stability of optical properties of TiN/ScN SLs. (a) Image showing that the TiN/ScN SL on F-mica after mechanical exfoliation can be gradually bent to a radius of ~ 2 mm without cracking. (b) Post-annealing and repeated bending tests for the SL2 sample. (c) The dielectric constants of SL2 extracted from SE measurements after each testing procedure.

**Table 1.** A comparison of different flexible multilayered HMMs in literature and this work. PVA, PMMA, PET, and PDMS stand for poly(vinyl alcohol), poly(methyl methacrylate), polyethylene terephthalate, and polydimethylsiloxane, respectively.[49-50]

|  | Flexible Au/PVA HMMs | Flexible Au/PMMA HMMs | Flexible TiN/ScN HMMs |
|---|---|---|---|
| Functional components | [Au(25 nm)/PVA(42 nm)]$_4$ and [Au(18 nm)/PVA(49 nm)]$_4$ multilayers | [Au(25 nm)/PMMA(30 nm)]$_4$ and [Au(25 nm)PMMA(40 nm)]$_4$ multilayers | [TiN(50 Å)/ScN(42 Å)]$_{10}$ and [TiN(50 Å)/ScN(76 Å)]$_8$ superlattices |
| Crystallinity | Non-single-crystalline | Non-single-crystalline | Single-crystalline |
| Substrates | PET | PDMS/Paper | F-mica |
| Operating temperature | Up to ~ 80 ℃ | Up to ~ 100 ℃ | Up to ~ 1000 ℃ |
| HMM ranges | > ~ 530 nm (Type-II) | > ~ 500 nm (Type-II) | 496 - 534 nm (Type-I) & > 641 nm (Type-II); 496 - 583 nm (Type-I) & > 750 nm (Type-II) |
| Bending radius | ~20 mm | ~1 mm | ~ 2 mm |